\def\mearth{{\rm\,M_\oplus}}
\documentclass[preprint,authoryear,12pt]{elsarticle}
\usepackage{amssymb}
\usepackage{aas_macros}

\begin{document}
 
\begin{frontmatter}

\title{Dynamical and collisional constraints on a stochastic late veneer on the terrestrial planets}

\author{Sean N. Raymond$^{a,b}$, Hilke E. Schlichting$^c$, Franck Hersant$^{a,b}$ \& Franck Selsis$^{a,b}$}
\address[1]{Univ. Bordeaux, LAB, UMR 5804, F-33270, Floirac, France; rayray.sean@gmail.com}
\address[2]{CNRS, LAB, UMR 5804, F-33270, Floirac, France}
\address[3]{Department of Earth and Space Sciences, University of California, Los Angeles, CA 90095, USA}

\begin{abstract}
Given their tendency to be incorporated into the core during differentiation, the highly-siderophile elements (HSEs) in Earth's mantle are thought to have been accreted as a ``late veneer'' after the end of the giant impact phase.  Bottke et al (2010) proposed that the large Earth-to-Moon HSE abundance ratio can be explained if the late veneer was characterized by large ($D = 1000-4000$~km) impactors.  Here we simulate the evolution of the terrestrial planets during a stochastic late veneer phase from the end of accretion until the start of the late heavy bombardment $\sim$500 Myr later.  We show that a late veneer population of $0.05 \mearth$ dominated by large ($D > 1000$~km) bodies naturally delivers a $\sim 0.01 \mearth$ veneer to Earth, consistent with geochemical constraints.  The eccentricities and inclinations of the terrestrial planets are excited by close encounters with the largest late veneer bodies.  We find the best agreement with their post-veneer orbits if either a) the terrestrial planets' pre-veneer angular momentum deficit $AMD_0$ was less than about half of the current one $AMD_{now}$, or b) $AMD_0 \le AMD_{now}$ and the veneer was limited to smaller ($D_{max} \le 2000$~km) bodies.  Veneer impacts on Venus, Earth and Mars were mostly accretionary but on Mercury and the Moon they were mostly erosive.  In $\sim 20\%$ of simulations an energetic impact occurred that could have removed $\ge 25\%$ of Mercury's mass, thereby increasing its iron mass fraction.  We show that, due to the erosive nature of larger impacts, the Moon cannot accrete any material from objects larger than 500-1000~km. The large Earth-to-Moon HSE abundance ratio is naturally explained if the late veneer included large impactors ($D \gtrsim 500-1000$km) regardless of their size distribution, as long as most of Earth's veneer came from large bodies.  The spin angular moment imparted by stochastic late veneer impacts was far in excess of the current ones for Mercury and Venus, meaning that their post-veneer spin rates were much faster.  The late veneer, if it included large impactors, was an accretionary phase for Venus, Earth and Mars but an erosive phase for Mercury and the Moon.  

\end{abstract}

\end{frontmatter}

\section{Introduction}

The so-called ``late veneer'' phase in Earth's accretion is sandwiched between two far more dynamic epochs.  The late veneer comes at the end of the giant impact phase of terrestrial accretion, which lasted 30-100 Myr~\citep{kleine09} and during which the growing Earth underwent numerous impacts with Moon- to Mars-sized planetary embryos~\citep{wetherill85,chambers98,agnor99,chambers01,raymond06b,raymond09c}.  Thus, the late veneer probably represents accretion during the final clearing-out of planetesimals leftover from terrestrial accretion, starting after the Moon-forming impact and dwindling in time for $\sim$500 Myr.  After the end of the late veneer came the late heavy bombardment, a spike in the impact flux that lasted a few hundred million years~\citep{tera74,wetherill75}.

Evidence for the late veneer comes from the existence of highly-siderophile elements (HSEs) in the mantles of Earth, Mars and the Moon.  Simply put, HSEs are ``iron-loving" elements that tend to partition into metal and should thus be removed from a planet's mantle during core formation.  The abundance of HSEs on Earth imply that it accreted at least an additional 0.3-0.7\% of an Earth mass of material with chondritic composition after the last core-forming event~\citep{kimura74,day07,walker09}, presumed to be the Moon-forming impact~\citep{benz86,canup01}.  The concentration of HSEs in the lunar mantle is 20-40 times lower than the concentration in Earth's mantle~\citep{day07}.  As the Earth's mantle is about 70 times more massive than the Moon's, this naively implies that the Earth accreted $\sim 1400-2800$ times more mass in HSEs during the late veneer than the Moon.  However, by taking into account the lunar crust as an additional, possibly dominant, repository for HSEs~\citep{walker04}, \cite{schlichting12} calculated a significantly higher total concentration of HSEs on the Moon.  \cite{schlichting12} found a Earth-to-Moon HSE abundance ratio -- meaning the ratio in the total masses, not concentrations, of HSEs in their mantles -- of 150-700~\citep[see also the discussion in][who favor a range from 200 to 1200]{bottke10}.  The large Earth/Moon HSE abundance ratio is problematic because the ratio of the physical cross sections of the Earth and the Moon is only 13.5.  The measured concentration of HSEs in the Martian mantle is similar to the Earth's~\citep{walker09,brandon12}. Taking into account the relative sizes of their mantles, this implies that Earth accreted roughly 9 times more mass in its late veneer than Mars.  

To date, two potential solutions to this problem have been proposed that make vastly different assumptions about the characteristic size of late veneer impactors.  ~\cite{bottke10} showed that the large Earth/Moon HSE abundance ratio can be reproduced by a veneer population skewed toward large objects such that almost all of the HSEs are delivered by a few big impacts with $D > 2000$~km.  The combination of the small number of impactors and the Earth's larger collisional cross section allow the Earth to accrete far more veneer mass than the Moon.  This model is consistent with the observed patchiness in Tungsten isotopes in Earth's mantle~\citep{willbold11}.  In contrast, the model of~\cite{schlichting12} requires a veneer made of very small ($D \sim 10$~m) planetesimals, which collisionally damp to very low eccentricities and thus increase the gravitational focusing factor of the Earth sufficiently to explain the Earth/Moon HSE abundance ratio. These small planetesimal could at the same time provide the required damping of the eccentricities and inclinations after giant impacts.  Bottke et al's model is consistent with planetesimals being born big~\citep{johansen07,morby09b} whereas Schlichting et al's model is consistent with collisional models that start from small planetesimals~\citep{weidenschilling11}.  Of course, all planetesimals must have undergone some collisional evolution by this point so it is unclear how the veneer size distribution relates to the initial one.

In this paper we explore the consequences of the ``stochastic late veneer''~\citep{bottke10} model for the dynamics and collisional history of the terrestrial planets.  We address the orbital excitation of the terrestrial planets, the collisional regime of veneer impacts on each planet, and the effect of the late veneer on planetary spins.  

The orbital excitation of the terrestrial planets can be quantified using the normalized angular momentum deficit $AMD$~\citep{laskar97}, which measures the difference in angular momentum of a set of orbits from coplanar, circular orbits with the same semimajor axes: 
\begin{equation} 
AMD = \frac{\sum_{j} m_j \sqrt{a_j} \left(1 - cos(i_j) \sqrt{1-e_j^2}\right)} {\sum_j m_j \sqrt{a_j}}, 
\end{equation}
\noindent where $a_j$, $e_j$, $i_j$, and $m_j$ refer to planet $j$'s semimajor axis, eccentricity, inclination with respect to a fiducial plane, and mass. The current $AMD$ of the Solar System's terrestrial planets is 0.0018; we refer to this value throughout the paper as $AMD_{now}$.  

Simulations of terrestrial planet formation have historically produced planetary systems that were overly dynamically excited, i.e., with $AMD > AMD_{now}$~\citep[e.g.][]{chambers98,agnor99,chambers01,raymond04}.  This problem was a result of numerical limitations.  Including a large population of planetesimals mediated this discrepancy and produced planetary systems with $AMD \sim AMD_{now}$~\citep{raymond06b,obrien06,raymond09c} because dynamical friction from the small bodies acts to damp random velocities in the larger ones.  However, some high-resolution simulations show large increases in the orbital excitation of the entire system at late times by objects that represent less than 1\% of the total system mass.  For example, in the simulation illustrated in Figures 2-12 of~\cite{raymond06b}, a series of close encounters with a roughly lunar-mass embryo more than doubled the eccentricity of the system's Earth analog, and roughly doubled the mass-weighted eccentricity of the entire terrestrial planet system~\citep[see in particular Figs. 8, 9 and 11 in][]{raymond06b}.  

The terrestrial planets' orbital eccentricities during the late veneer epoch are poorly constrained.  A large-scale re-arrangement of the giant planets' orbits is thought to have occurred during the late heavy bombardment~\citep[the so-called ``Nice model'';][]{tsiganis05,morby10}.  During this instability, secular resonances with the giant planets may have crossed the terrestrial planet region.  If the orbits of Jupiter and Saturn evolved slowly the terrestrial planets' eccentricities and inclinations would have been excited to larger than the current values~\citep{brasser09,agnor12}.  However, scenarios that invoke rapid evolution of the giant planets' orbits adequately reproduce the architecture of both the giant and terrestrial planets~\citep{brasser09,morby09a}. \cite{brasser13} showed that if the terrestrial planets' $AMD$ at early times was larger than about $0.7 \, AMD_{now}$, then their eccentricities would have been over-excited by the giant planet instability.  Given that chaotic diffusion tends to increase rather than decrease orbital excitation in time~\citep{laskar93}, the only reasonable constraint that we can place is that the excitation of the terrestrial planets after the late veneer phase could not have been larger than this.


\section{Simulations}

Our simulations start immediately after the end of the giant impact phase of terrestrial accretion.  At this stage there remained a substantial population of planetesimals too small to cause another core-forming impact on Earth.   \cite{bottke10} argued that objects for which the diameter of the iron core was larger than the depth of the Earth's presumed magma ocean would trigger a differentiation event~\citep{rubie03} and thus constrained the maximum size of the veneer to have diameters $D<2000-4000$km.

We model this late veneer population by assuming that it follows a power law in size:
\begin{equation}
\frac{{\rm d}N}{{\rm d}D} \propto D^{-q},
\end{equation} 
where ${\rm d}N$ is the differential number of objects in a bin of width ${\rm d}D$ and $q$ is the size-distribution exponent.  ~\cite{bottke10} found that the Earth/Moon HSE abundance ratio was best reproduced for $q=1-2$ and maximum impactor size $D_{max}$ = 3000-4000~km.  We adopt the same parameter range with $q$ = 1, 1.5 and 2 and $D_{max}$ = 2000, 3000 and 4000~km, for a total mass in the veneer population of $0.05-0.1 \mearth$.

To generate a population of late veneer impactors, we determined the size of each object by drawing from a size-frequency distribution with a fixed $q$ value and $D$ between 1000-4000 km.  We neglected objects smaller than 1000 km because 1) they are too numerous to include a reasonable number of objects with an N-body approach, and 2) for $q \leq2$ the total veneer mass is dominated by the largest objects such that the $D<1000$~km population represents only about 1/4 of the total veneer mass.  Although their dynamical influence is probably small, these small bodies may be very important in determining the Earth-to-Moon HSE abundance ratio, as we discuss in detail in Sections 7 and 8. 

We assigned an orbit to each veneer particle, consistent with recent simulations of terrestrial planet formation, in particular~\cite{raymond06b} and~\cite{walsh11}.  The semimajor axis was randomly chosen between 0.5 and 1.7 AU, the eccentricity between 0.1 and 0.6, and the inclination between zero and $20^\circ$.  The other orbital angles were chosen at random from $0-360^\circ$.  The physical density of veneer impactors was fixed at $3 \, {\rm g \, cm^{-3}}$.  

Given that the orbits of the terrestrial planets during this phase are not known, we chose four different starting configurations with a range of initial $AMD$ values.  In the first case, the terrestrial planets were placed on orbits with their current semimajor axes but with zero eccentricity and inclinations of $0.1-0.2^\circ$ for a starting $AMD$ of less than $10^{-6}$.  This isolates the eccentricity excitation by the late veneer population.  In the next two cases we chose $AMD_0$ = 0.5 and 1 times the current value $AMD_{now}$.  In the case of $AMD_0 = 0.5 \, AMD_{now}$, the actual (not normalized) $AMD$ of each planet's orbit was simply halved, then corresponding values of $e$ and $i$ were chosen for each planet, taking care not to exceed any of the current values.  For $AMD_0 = AMD_{now}$ we simply put the terrestrial planets' on their current orbits.  For the final starting configuration we gave the terrestrial planets their current orbital elements, then artificially increased the eccentricity of each planet to 0.1.  This produced an $AMD_0$ value of 3.24, which we refer to as $AMD_{big}$.  

The different choices of $AMD_0$ reflect our uncertainty in the orbits of the terrestrial planets immediately after the giant impact phase.  For giant impacts to occur the planets' orbits must have been at least somewhat eccentric, simply to cause their orbits to cross.  However, the $AMD$ of the terrestrial planet {\it system} need not be very excited for a given pairwise impact to occur, especially if the impactor is significantly less massive than the target.  From an inspection of accretion simulations that adequately reproduce the terrestrial planets, the chosen values appear to reflect the range of reasonable $AMD_0$ values.  

In all cases the terrestrial planets were given their current masses and physical densities.  Jupiter and Saturn were also included in the simulations in a low-eccentricity, nearly-coplanar 3:2 resonant configuration with Jupiter at 5.4 AU and Saturn at 7.2 AU.  These orbits that are consistent with both their prior migration in the gaseous disk~\citep{masset01,walsh11,pierens11} and their later evolution during the LHB instability~\citep{morby07b,batygin10,nesvorny12}.  

Each system was integrated for 500 Myr using the hybrid {\tt Mercury} integrator \citep{chambers99} with a 6 day timestep.  Objects were removed if their orbital distances became smaller than 0.2 AU~\citep[assumed to collide with the Sun; see][for numerical tests of energy conservation as a function of minimum perihelion distance]{raymond11} or exceeded 100 AU (assumed to be ejected from the planetary system).  Collisions were treated as inelastic mergers conserving linear momentum (but see Section 3).  The terrestrial and giant planets were treated as fully-gravitating bodies.  Veneer particles felt the gravity of the planets but, to reduce the computational cost, did not self-gravitate.  In terms of conservation of energy $E$ and angular momentum $L$, for each simulation ${\rm d}E/E < 2 \times 10^{-6}$ and ${\rm d}L/L < 2 \times 10^{-11}$.  

For each set of parameters we ran eight simulations, for a total of 326 simulations.\footnote{We only performed simulations with $M_{tot} = 0.1 \mearth$ for $D_{max} = 4000$km with $AMD_0 =$0 and $0.5\, AMD_{now}$.  For $D_{max} = 2000$km we only performed 24 simulations, with $q=2$.}  In some simulations, Mercury's eccentricity was increased to values high enough that it impacted the Sun; given that the Sun's radius is 0.2 AU in the simulations this implies $e_{Mercury} > 0.48$.  This occurred in 12 simulations with $D_{max} = 4000$km: 2 for $AMD_0 \le 0.5 \, AMD_{now}$, and 10 for $AMD_0 = AMD_{max}$.  It did not occur in any simulations with $D_{max} \le 3000$km.  Simulations in which Mercury was destroyed were removed from the rest of the analysis.  

We assumed that veneer particles were chondritic in their HSE compositions and not fragments of differentiated bodies.  

\section{Collisional model}

In our simulations each collision was treated as an inelastic merger.  In reality, many collisions should not have resulted in perfect merging.  To account for imperfect accretion, we post-processed each simulation after it was run to determine what the outcome of each collision should have been.  We extracted the speed and angle of each impact from {\tt Mercury} runfiles using a leapfrog integrator to reach the moment of collision.  

We applied the collisional model of \cite{leinhardt12}, who mapped out the parameter space of gravity-dominated impacts.  Each collision is characterized by the two objects' masses and radii, the impact angle and the impact speed.  The impact speed is expressed in terms of the two-body escape speed 
\begin{equation}
V_{esc} = \sqrt{\frac{2 \, G\, (M_p+M_t)}{R_p+R_t}}, 
\end{equation}
where $G$ is the gravitational constant, $M$ and $R$ are the two objects' masses and radii, and subscripts $p$ and $t$ refer to the projectile and target, respectively.  

There are three common outcomes to the collisions in our simulations.  First, accretionary impacts are relatively low-speed and close to head-on.  They lead to net growth.  Second, "hit and run" impacts occur at grazing angles such that the two bodies effectively bounce off of one another with no net mass transfer~\citep{asphaug06,asphaug10}.\footnote{There exists a tiny slice in the parameter space of hit-and-run impacts in which enough energy is dissipated for the projectile to be gravitationally captured by the target, called ``graze and merge'' by~\cite{genda12}.  Not a single impact in our 326 simulations fell into that category.  In contrast, ~v\cite{chambers13} found such impacts to be relatively common in impacts between similar-sized embryos in accretion simulations.  Indeed, the accretion phase appears more conducive to graze-and-merge events than the late veneer phase because of the different impact distributions during these phases, in particular the regimes of mass ratio and relative impact velocity.} Third, high-speed impacts are so energetic that they lead to net erosion of the largest body.  We quantified these outcomes with the mass of the largest remnant body following \cite{leinhardt12}~\citep[see also][]{genda12}.  Curves dividing the different outcomes are shown in the bottom right panel of Fig.~\ref{fig:vimp_acc}.

In non-accretionary collisions, some escaping fragments may come from the target and not just the partially-accreted projectile.  The balance between the sources of fragments should affect the amount of HSEs delivered in a given impact.  There is little to no information on this balance from studies of gravity-dominated collisions, so we make the simple assumption that the fragments are all derived from the projectile, meaning that the amount of HSE delivered per impact is underestimated (although we don't know by how much).  In practice, we are not strongly affected by this assumption because veneer impacts on the Earth are dominated by near-perfect mergers and hit-and-run impacts, neither of which are affected by this assumption.  We also note that our post-processing of the simulations does not allow us to track the evolution of fragments in imperfect mergers.  Fragments are assumed to have been lost from the system.  

\section{Simulation results: dynamics}
\begin{figure}
\includegraphics[width=0.95\textwidth]{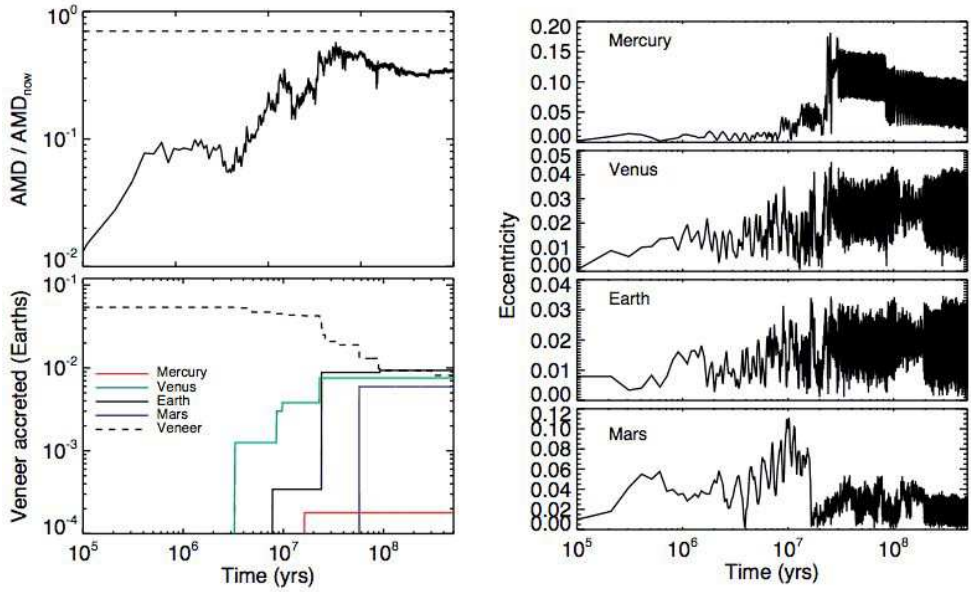}
\caption{Evolution of a simulation in which the $AMD$ of the terrestrial planets starts at zero and is excited to roughly 1/3 of its current value $AMD_{now}$.  {\bf Top left:} The terrestrial planets' $AMD$ as a function of time.  {\bf Bottom left:} Mass evolution of the system: the dashed line represents the surviving mass in veneer particles, and the solid curves represent the veneer mass accreted by Earth (black), Venus (green), Mars (blue), and Mercury (red).  This includes the effect of imperfect accretion described in Section 3.  {\bf Right:} Terrestrial planet eccentricities vs. time.}
\label{fig:evol1}
\end{figure}

Figure~\ref{fig:evol1} shows the evolution of a simulation that is consistent with the Solar System's evolution.  In this case $AMD_0 = 0$, $q = 1.5$, $D_{max}$ = 4000~km, and $M_{tot} = 0.05 \mearth$ in 20 veneer particles.  In the first ten million years, the terrestrial planets' $AMD$ increased from nearly zero to roughly half the current value.  After that point the $AMD$ remained relatively stable with values of $0.3-0.6 AMD_{now}$.  The evolution of the planetary eccentricities in the example simulation generally mirrors that of the $AMD$ but each planet's evolution is unique because it is determined by stochastic close encounters with a small number of objects.  Over the course of the simulation, the maximum eccentricities reached by each terrestrial planet was somewhat smaller than its maximum value from secular oscillations in the current Solar System architecture~\citep{quinn91}.  The maximum eccentricity varied from simulation to simulation and as a function of the parameters we tested, but were consistently close to the current ones.

The excitation of the terrestrial planets' orbits occurs via a combination of close encounters with massive veneer particles and secular perturbations from the other planets.  A veneer particle excites a planet's eccentricity by either a direct perturbation during a close encounter or by a perturbation to the eccentricity of a different planet that is transmitted to the other planets via secular coupling.  In the absence of veneer particles, the eccentricities of the $AMD_0 \approx 0$ terrestrial planets oscillate with amplitudes of just a few$\times 10^{-3}$.  However, close encounters with large veneer particles cause the terrestrial planets' evolution to diverge from a veneer-free system in just a few thousand years and to systematically reach eccentricities of $\sim 0.01$ or higher.  

Each of the terrestrial planets accreted a late veneer in the example simulation (Fig.~\ref{fig:evol1}).  Earth accreted $9 \times 10^{-3} \mearth$ from three veneer particles in the simulation from Fig.~\ref{fig:evol1}.  Two impactors were on the small side of our simulated spectrum ($D$ = 1082 and 1270~km) but one was roughly Moon-sized ($D$ = 3180~km).  Venus accreted $7 \times 10^{-3} \mearth$ in 4 impacts, each with $D$ between 1500 and 2500~km.  Mars accreted a single large veneer object with $D$ = 2825~km.  Mercury suffered a single, modestly off-center ($\sin{\theta} = 0.71$), relatively high-speed ($19.1 \, km \, s^{-1}$ = 5.4 times Mercury's escape velocity) veneer impact with a $D$ = 1058~km object.  The impact was net accretionary, and Mercury retained about half of the impactor mass, $1.7 \times 10^{-4} \mearth$.  Of the 20 veneer particles, 9 collided with the terrestrial planets, 5 were ejected after close encounters with Jupiter, 3 collided with the Sun, and 2 survived on orbits exterior to that of Mars.

\begin{figure}
\includegraphics[width=\textwidth]{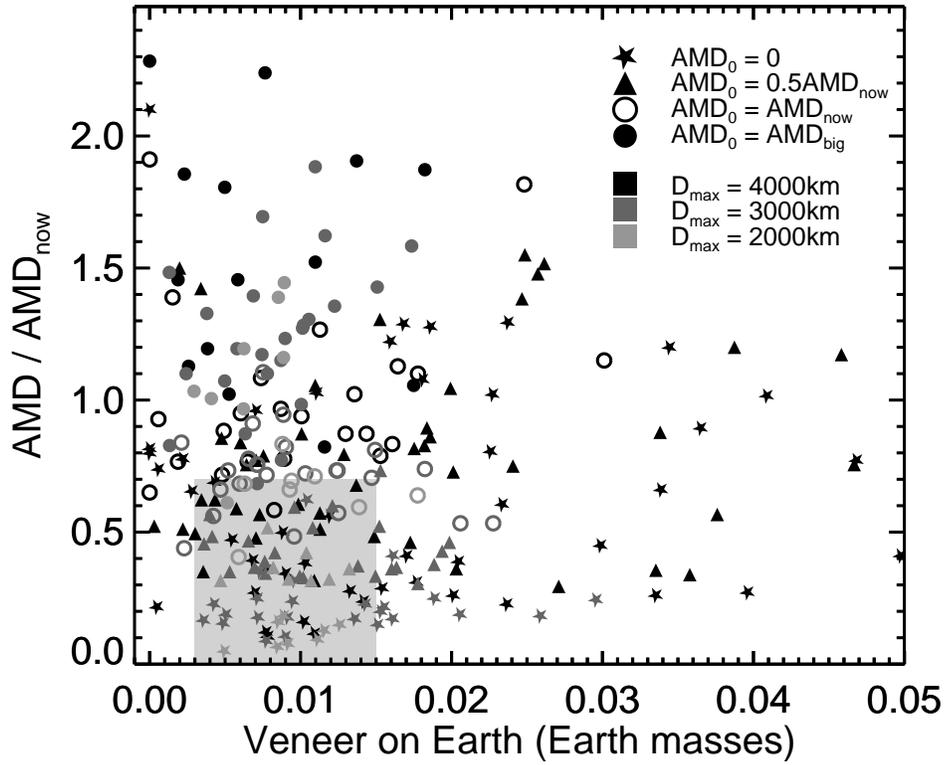}
\caption{The angular momentum deficit of the terrestrial planets $AMD$ relative to their current value $AMD_{now}$ as a function of the veneer mass accreted by Earth.  The black symbols represent simulations with $D_{max}=$ 4000~km and the grey symbols simulations with $D_{max} =$ 3000~km.  The different symbols refer to simulations with different values of $AMD_0$.  The shaded region represents outcomes that are consistent with the present-day Solar System. }
\label{fig:amd-veneer}
\end{figure}

Figure~\ref{fig:amd-veneer} shows the final $AMD$ of all of our simulations as a function of the total veneer mass accreted by Earth.  There is no trend between the Earth's veneer, which ranges from zero to almost $0.05 \mearth$, and the $AMD$, which ranges from about 0.1 to 2.3 times the current value $AMD_{now}$.  However, there is clear clustering of sets of simulations with different parameters.   

Given the observational and dynamical constraints discussed above, a successful outcome must satisfy two constraints.  First, the Earth's late veneer must be in the range $0.003-0.015 \mearth$.  This range is constrained by the concentration of HSEs in Earth's mantle~\citep{walker09}, extending from the inferred lower limit to roughly twice the inferred upper limit to allow for some loss.  Second, the $AMD$ of the terrestrial planet system must be less than $0.7$ times the current value to remain consistent with dynamical excitation during the LHB~\citep{brasser13}. 

We evaluate the fraction of simulations with successful outcomes for different parameters.  Fixing $M_{tot} = 0.05 \mearth$, the success rate is higher for smaller $D_{max}$ and for $AMD_0 \le 0.5 \, AMD_{now}$.  For $D_{max} = $4000~km the success rate was 55\% (13/23)for $AMD_0=0$, 35\% (8/23) for  $AMD_0=0.5 \, AMD_{now}$ and just 4-7\% (1/24 and 1/14, respectively) for  $AMD_0=AMD_{now}$ and $AMD_0=AMD_{big}$.  For $D_{max} =$ 3000~km the success rate was $\sim65$\% for $AMD_0 \le 0.5 \, AMD_{now}$ (15/24 and 16/24) but just 21\% (5/24) for $AMD_0=AMD_{now}$ and 4\% (1/24) for $AMD_0=AMD_{big}$.  For $D_{max} =$ 2000~km the success rate was 100\% for $AMD_0 \le 0.5 \, AMD_{now}$ (16/16), 62.5\% (5/8) for $AMD_0=AMD_{now}$ and 12.5\% (1/8) for $AMD_0=AMD_{big}$.


For smaller $q$ the veneer is dominated by progressively larger bodies, so for a fixed veneer mass the number of veneer particles decreases with decreasing $q$.  We found modest differences in the excitation of the terrestrial planets' $AMD$ for veneer populations with different $q$ values.  For simulations with $AMD_0 = 0$ and $D_{max} = 4000$~km, the median $AMD$ is higher for $q=1$ than for $q=1.5$ or 2, with a median of $0.8 \, AMD_{now}$ for $q=1$ compared with $0.39 \, AMD_{now}$ for $q = 1.5$ or 2.  This difference is barely statistically significant for $AMD_0 = 0$ and $D_{max} = 4000$~km, with a $p$ value of 0.01 from a K-S test.  The difference is marginally significant for $D_{max} =$ 3000~km, with $p=0.09$.  However, the $AMD - q$ trend disappears among systems with larger $AMD_0$ for $D_{max} =$ 3000~km or 4000~km.  

The final $AMD$ of the terrestrial planets is mainly determined by $D_{max}$ and $AMD_0$. The $AMD$ is significantly more excited in simulations with larger $D_{max}$, suggesting that it is the few most massive veneer particles that dominate the excitation.  Indeed, for each set of simulations with fixed $q$ and $AMD_0$ there is a strongly statistically significant increase in the $AMD$ for the simulations with $D_{max}$ = 4000~km compared with $D_{max}$ = 3000~km or 2000~km.   A Kolmogorov-Smirnov test yielded probabilities $p$ of less than $10^{-4}$ that the $AMD$ values for $D_{max} = 4000~km$ were drawn from the same distribution as for $D_{max}$ = 3000~km or 2000~km.  When comparing simulations with $D_{max}$ = 2000~km vs. 3000~km, there is only a statistically significant lower $AMD$ for $D_{max}$ = 2000~km for the case of $AMD_0 = 0$ ($p = 1.4 \times 10^{-3}$).  For higher $AMD_0$ the final $AMD$ is modestly lower for $D_{max}$ = 2000~km but it is not statistically significant.

The terrestrial planets' final $AMD$ values remain somewhat segregated by their initial values but with substantial overlap (see Fig.~\ref{fig:amd-veneer}).  For $D_{max}$ = 4000~km and $M_{tot} = 0.05 \mearth$, the median final $AMD$ values for $AMD_0$ = 0, $0.5\, AMD_{now}$, $AMD_{now}$, and $AMD_{big}$ was $0.54 \, AMD_{now}$ were 0.38, 0.68, 0.94 and $1.5 \, AMD_{now}$, respectively.  For $D_{max} = 3000$~km the corresponding final $AMD$ values were 0.2, 0.42, 0.73, and 1.3 $\, AMD_{now}$.  For $D_{max} = 2000$~km the corresponding final $AMD$ values were 0.13, 0.36, 0.68, and 1.2 $\, AMD_{now}$. 

This suggests that the veneer systematically excites the terrestrial planets' $AMD$ to a level that depends on $D_{max}$ but only excites the $AMD$ beyond this value in a stochastic way.  For $D_{max}$ = 4000~km the level of systematic excitation is on the order of half of the current $AMD$ but for $D_{max}$ = 3000~km this level is closer to 0.2 times current and for $D_{max}$ = 2000~km it is about 0.1 times current.  When the terrestrial planets' $AMD$ is larger than this critical value, the net effect of the veneer is to damp rather than excite eccentricities.  Indeed, for all sets of simulations with $AMD_0 = AMD_{now}$ or $AMD_{big}$ the final median $AMD$ was smaller than the starting $AMD$, and the veneer-induced decrease in $AMD$ was larger in systems with more particles, i.e. for smaller $D_{max}$.  

The simulations with $D_{max} = 2000$~km were the most successful in maintaining a small $AMD$.  Starting with $AMD_0 \le AMD_{now}$, the vast majority of simulations satisfied our constraints on the final simulation $AMD$ be less than $0.7 \, AMD_{now}$.  This criterion were met in simulations with larger $D_{max}$ but for a more restricted range of initial $AMD$: $AMD_0 \approx 0$ for $D_{max} = 4000$~km and $AMD_0 \le 0.5 \, AMD_{now}$ for $D_{max} = 3000$~km.  

\section{Simulation results: collisions}
We found no statistically significant trend between the $AMD$ and $M_{tot}$.  Given that our range in $D$ was fixed, a higher $M_{tot}$ simply implies a larger number of veneer particles.  This increased the number of close encounters such that the number of strong perturbations was increased, but also enhances the damping, swarm-like behavior of the population.  

\begin{figure}
\includegraphics[width=\textwidth]{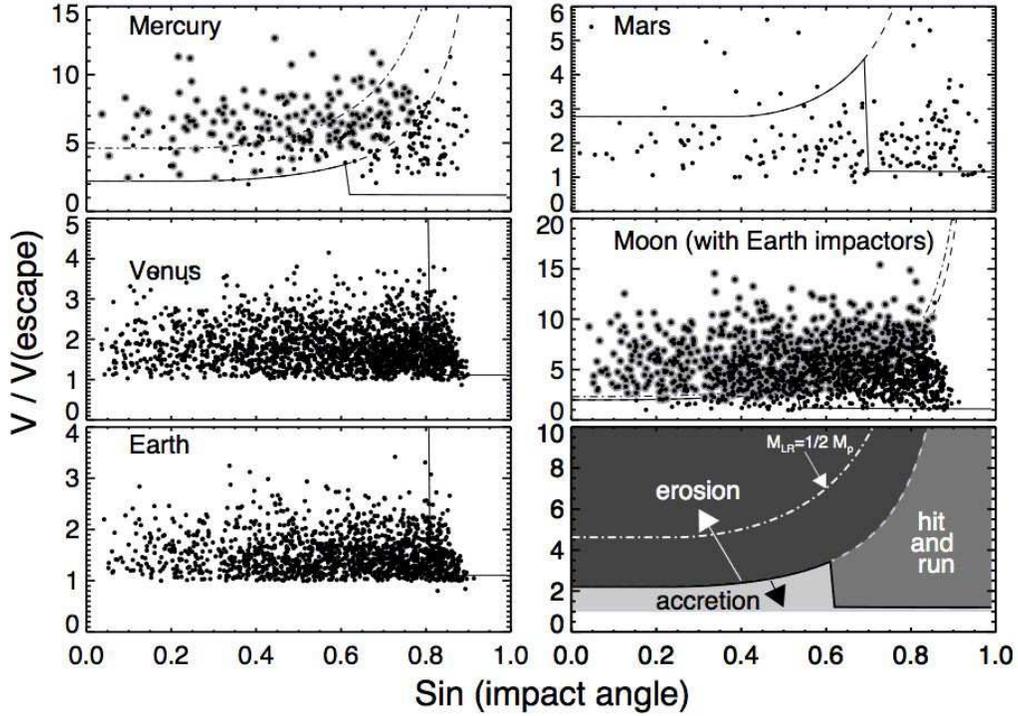}
\caption{Impact velocity (expressed as a fraction of the two-body escape speed) as a function of the impact parameter $b$, i.e. the sine of the impact angle for all of the collisions on each terrestrial planet in our simulations.  Curves that determine the accretion outcome are overplotted in each panel assuming a $D=3000$~km projectile ~\citep[see Section 3 and ][]{leinhardt12}.  The bottom right panel explains the layout: below the solid curves  impacts are accretionary; this is the lightly shaded area in the bottom right panel.  Hit and run impacts are at large impact angles, between the solid and dashed curves (medium shaded area in bottom right panel).  Above the dashed curves impacts are erosive (dark shaded area), and the dot-dashed line (only visible for Mercury and the Moon) shows where the largest post-impact remnant contains half of the mass in the planet.  For the Moon and Mercury, the gray shaded impacts were erosive.}
\label{fig:vimp_acc}
\end{figure}

Figure~\ref{fig:vimp_acc} shows the velocity and impact angle of each impact that occurred in our simulations. For Earth, Venus and Mars the typical veneer impact occurred at modest velocity, with a median $v/v_{esc}$ of 1.45 for Earth, 1.76 for Venus, and 1.9 for Mars.  Every impact on these three planets was in either the accretionary or hit-and-run regimes.  

Among simulations with $M_{tot} = 0.05 \mearth$, Earth accreted zero (in three cases) to $0.038 \mearth$ with a median of $9 \times 10^{-3} \mearth$.  Given that this is consistent with the geochemical constraints~\citep{walker09}, a total mass of $\sim 0.05 \mearth$ in the initial veneer population appears to be realistic.  In the same subset of simulations Venus accreted between zero (in one case) and $0.032 \mearth$ with a median of $0.011 \mearth$.  

Mars was hit by a veneer particle in about half (102 of 224 = 46\%) of the simulations.  In 25 of those cases the impacts were purely hit-and-run and resulted in no net accretion. In accretionary simulations Mars' late veneer was as high as $0.011 \mearth$ and the median for simulations with at least one [accretionary] impact was $7.2 \times 10^{-4}$ [$9.9 \times 10^{-4}$] $\mearth$.  In cases when Mars did accrete a veneer the median Earth/Mars veneer mass ratio was 8, consistent with the Earth/Mars mantle volume ratio of $\sim 9$ given that the concentration of HSEs in the mantles of each are comparable~\citep{walker09}.  

Impacts on Mercury occurred at a much higher fraction of the escape speed.  Mercury suffered at least one impact in 131 of 224 (58\%) simulations with $M_{tot} = 0.05 \mearth$.  The median $v/v_{esc}$ for veneer impactors was 5.7, and more than 10\% of impacts occurred at $v/v_{esc} > 8$.  Mercury's impactors tended to come from somewhat exterior to Mercury's orbit -- with semimajor axes typically between 0.5 and 0.7 AU -- on eccentric orbits, thus resulting in large collision speeds.  The dividing line between net accretion and erosion for large (3000~km) veneer impactors on Mercury is at $v/v_{esc} \gtrsim 2$ (Fig. 3) depending on the impact angle.  Among all veneer impacts on Mercury, only 31\% resulted in net growth; 19\% were hit-and-run impacts, and the remaining 50\% caused net erosion of the planet (see Fig. 3).  Among the 148 total erosive impacts, most (83\%) removed less than 25\% of Mercury's initial mass.  However, we calculated the mass of the largest post-impact remnant to be less than 75/25/10\% in 25 (17\%) / 11 (7\%) / 4 (3\%) of the total erosive impacts.  The most erosive impacts tended to be with relatively large impactors ($D \ge 2500$km), high speed and close to head-on.  Among the net-erosive simulations, Mercury was eroded by a median of 4\% of its current mass, for a net median decrease of $0.002 \mearth$.  Among all simulations with impacts, the median outcome was erosion of $\sim$1\% of Mercury's current mass.  

Given Mercury's high iron content, it has been speculated that a giant impact late in the accretion phase removed a large portion of the planet's mantle~\citep{benz88,benz07}.  Our simulations show that there is a chance that such an impact could have occurred during the late veneer phase.  In that case, the large velocity of the impactor would be a direct consequence of eccentricity excitation driven by the other terrestrial planets.  However, we note that the current-day Mercury would be represented by the {\em end product} of our simulated impacts, meaning that the simulated Mercury would have had to start as a much larger body.  An open question is whether the fragments from a mantle-stripping impact would be re-accreted, dispersed throughout the inner Solar System or ground to dust~\citep[see, for example,][with regards to the evolution of debris from the Moon-forming impact]{jackson12}. 

\section{Consequences for planetary spins}
The primordial rotation of planetary embryos is systematically prograde from the accretion of swarms of small bodies~\citep{schlichting07,johansen10}.  However, the giant impact phase tends to randomize the planets' obliquities~\citep{kokubo07} and the spin rate is likely determined by the last few giant impacts~\citep{dones93,agnor99}.  

Stochastic late veneer impacts can have a pronounced effect on the planets' spins.  Although the total mass in the late veneer is much smaller than the planets' masses, $D=1000-4000$~km objects impart a strong impulse.  Grazing impacts -- that would naively impart the strongest spin -- do not accrete (see Fig. 3) and therefore we assume that they have no effect on the planets' spins.  However, even a modest impact angle can impart a strong spin for a high-speed, high-mass impact.  

\begin{figure}
\includegraphics[width=\textwidth]{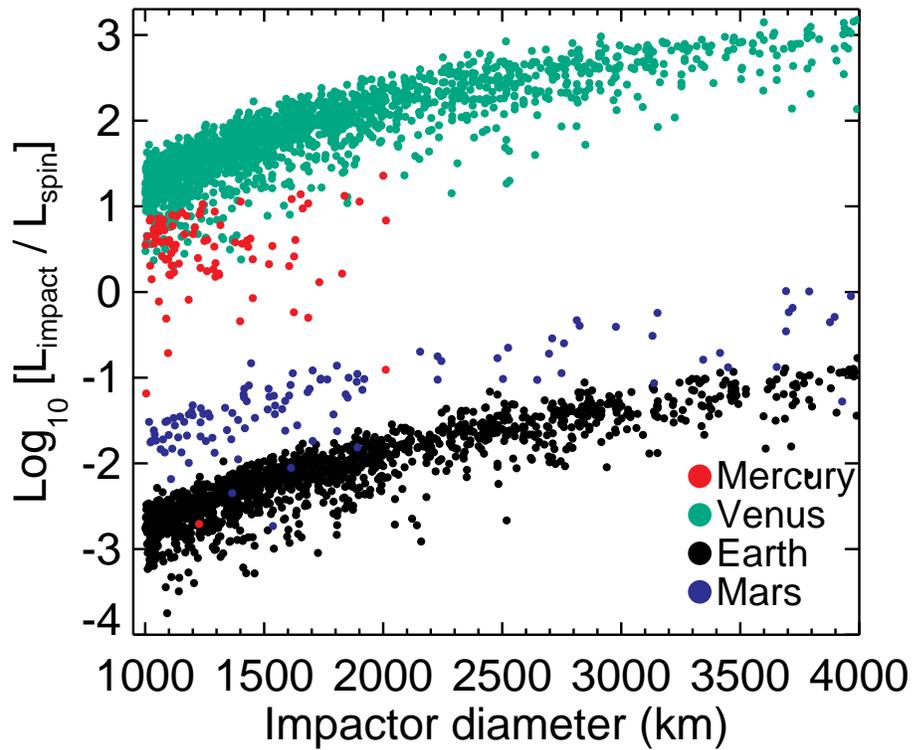}
\caption{Spin angular momentum delivered to the terrestrial planets in each late veneer impact as a function of the impactor's size, normalized by each planet's present-day spin angular momentum.  We have assumed that $L_{spin}$ scales linearly with the mass actually accreted during the impact as well as a 10\% efficiency for the conversion from impact- to spin angular momentum~\citep{love97}. }
\label{fig:limp_d}
\end{figure}

Figure~\ref{fig:limp_d} shows the expected spin angular momentum $L_{spin}$ delivered by each accretionary veneer impact, neglecting hit-and-run and erosive  impacts.  The amount of angular momentum delivered per impact is scaled by the mass that is actually accreted by the planet according to our collisional model.  This value is subsequently multiplied by 0.1 to account for additional loss of angular momentum during partial accretion, as fragments remove proportionally more specific angular momentum than accreted mass~\citep{love97}.  

The spin angular momentum delivered to Venus and Mercury by late veneer impacts is far larger than their current spin angular momenta.  This is largely a consequence of their current low spin rates.  Indeed, tidal dissipation is thought to have significantly altered the spin rates of both planets since their formation~\citep{correia01,correia09}.  If they underwent an early giant impact and/or stochastic late veneer phase, their primordial spin rates were necessarily orders of magnitude faster than their current ones.  

In contrast, the spin rates of Earth and Mars are only modestly affected by the late veneer.  No impact imparted more than 16\% of the Earth's current spin angular momentum.  Simulated impacts on Mars delivered up to 1.03 times its current spin angular momentum, but less than 10\% of veneer impacts on Mars delivered more than half of its current spin angular momentum.  

Assuming a zero initial rotation, the obliquities of our simulated terrestrial planets with late veneers are isotropically distributed.  As was the case for the spin rate, we do not expect the late veneer to have a strong influence on the obliquities of Earth or Mars.  However, it is plausible that the late veneer impactors could have imparted a retrograde rotation on the inner planets, especially if Mercury and Venus' primordial spins were relatively slow.  Although we have little direct information about their primordial rotation states, the current peculiar spin states of both Venus and Mercury can be most consistently reproduced if both planets' initial rotation was retrograde~\citep{correia01,wieczorek12}.  Indeed, we find that Venus' and Mercury's final rotation states are retrograde in almost exactly half of the simulations, although their outcomes do not correlate so they are both retrograde only about one quarter of the time.  

\section{The Earth/Moon HSE abundance ratio}
Although we did not include the Moon in our simulations, its impact environment would be similar to Earth's.  The impact velocity $v_{imp}$ is related to the random velocities of impactors $v_{rand}$ as 
\begin{equation}
v_{imp}^2 = v_{esc}^2 + v_{rand}^2,
\end{equation} 
so we can use the velocities of simulated impacts on Earth to calculate $v_{rand}$.  We then assume that the Moon was impacted by a population of bodies with the same $v_{rand}$ and, for convenience, with the same impact angles and masses as on Earth.  We must also take into account that impacts on the Moon are accelerated by the Earth's gravity.   The impact speed on the Moon $v_{imp, Moon}$ taking this into account is:
\begin{equation}
v_{imp,Moon}^2 = v_{rand}^2 + v_{esc,Moon}^2 + v_{esc,Earth-Moon}^2, 
\end{equation}
where $ v_{esc,Moon}$ is the escape speed from the Moon of 2.4 $km \, s^{-1}$, $v_{esc,Earth-Moon}$ is the escape speed from the Earth at the Moon's orbital distance $a_{Earth-Moon}$. Given that the Moon is receding from the Earth due to tidal interactions~\citep{kaula64,goldreich66}, $a_{Earth-Moon}$ was smaller in the past and the Earth's escape speed at $a_{Earth-Moon}$ was correspondingly higher, as high as 3.5 $km \, s^{-1}$ for $a_{Earth-Moon} = 10 R_\oplus$ or even 5 $km \, s^{-1}$ for $a_{Earth-Moon} = 5 R_\oplus$, compared with the current value of 1.4 $km \, s^{-1}$.  As the Moon receded from the Earth, the effect of Earth's gravitation on the impact speed decreased.  During the late veneer phase the Earth-Moon distance was evolving, although the early phase of evolution was the fastest and the Moon was probably beyond $\sim 40 R_\oplus$ within 200 Myr~\citep{schlichting12}.  Given that the veneer impacts in our simulations are skewed to earlier timescales, we used an intermediate value of 2.4 $km \, s^{-1}$ in our calculation of $v_{imp, Moon}$, coincidentally the same value as $v_{esc,Moon}$.  

This allows us to derive a reasonable population of impactors on the Moon.  This population is dominated by high-velocity impacts, with a median $v/v_{esc}$ of 5.07 and 4\% of impacts with $v/v_{esc} > 10$.  Most of these impacts -- 68\% -- are in the erosive regime: in 38\% / 27 \% / 16\% of cases the Moon would have lost more than 25\% / 75\% / 90\% of its current mass.  An additional 26\% of impacts are in the hit-and-run regime and only the remaining 6\% are accretionary.  

The analysis in the previous paragraph assumed that the same-sized objects impacted Earth and the Moon.  In contrast, \cite{bottke10} assumed that this could not be the case because if so, the Earth-to-Moon accretion ratio should simply be the ratio of their gravitationally-enhanced collisional cross sections $C$~\citep{safronov69}:
\begin{equation}
C = \pi R^2 \left(1+\frac{v_{esc}^2}{v_{rand}^2}\right),
\end{equation}
\noindent where $R$ is the accreting body's radius and $v_{esc}$ its escape speed.  From our simulations, $v_{rand}$ for veneer impactors on Earth ranged from 5.6 to 33 $km \, s^{-1}$ with a median of $11.7 \, km \, s^{-1}$.  Given that the random velocities are far larger than the Moon's escape speed of $2.4 \, km \, s^{-1}$, the Moon's cross section is simply its geometric one (i.e., $C_{Moon} \approx \pi R_{Moon}^2$).  Since its escape velocity of $11.2 \, km \, s^{-1}$ is comparable to the typical $v_{rand}$, the Earth's cross section is modestly gravitationally-enhanced.  For the median value of $v_{rand}$, the Earth-to-Moon collisional cross section ratio is 25.  The ratio in their total veneer masses, as interpreted from HSE concentrations in their mantles, is between 150 and 2800 (see discussion in Section 1).  With a shallow size distribution extending to large sizes and a 25-fold higher probability of impacting Earth, \cite{bottke10} showed that even assuming perfect accretion the large HSE abundance ratio could be reproduced with a significant probability thanks essentially to small number statistics.  

What we have shown is that, assuming the same impactor masses, all collisions (except for the 3\% of hit-and-run impacts) on Earth result in net accretion whereas 93\% of the same collisions result in no net accretion -- or, in most cases, in net erosion -- on the Moon.  In this context the Earth was able to accrete HSEs from large impactors whereas the Moon could not. 

\begin{figure}
\includegraphics[width=\textwidth]{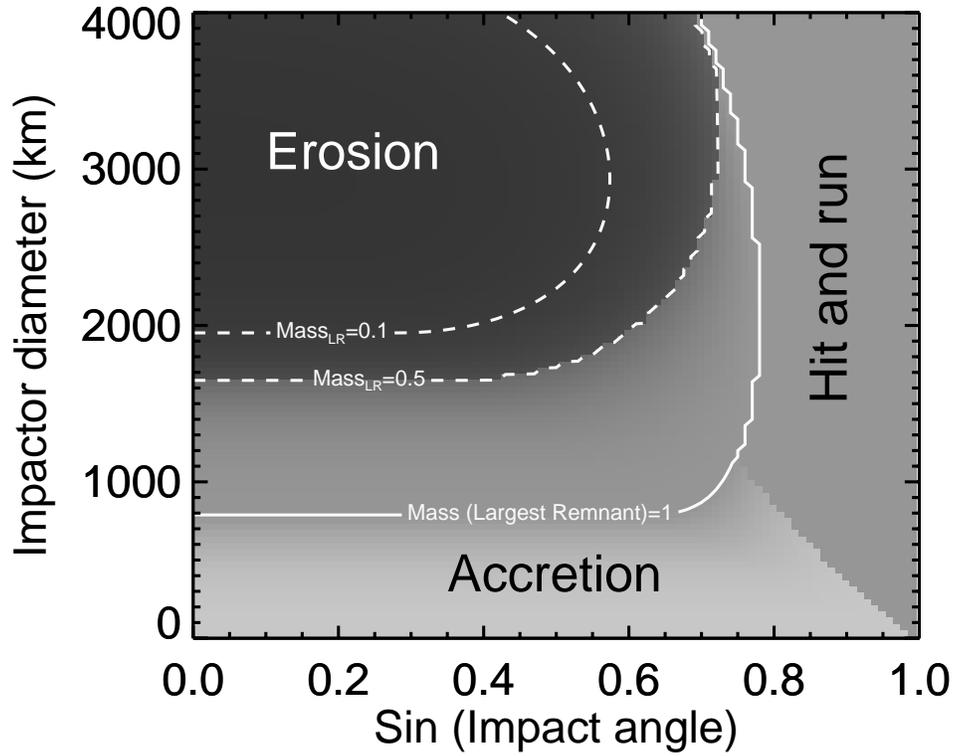}
\caption{Outcome of collisions with the Moon for a range of impactor sizes and impact angles, all assuming $v_{rand} = 11.7 \, km \, s^{-1}$, derived from the median impact velocity in the simulations.  The colors correspond to the mass of the largest post-impact remnant relative to the Moon's current mass, and the contours of the remnant mass are shown for values of 0.1, 0.5 and 1 lunar masses.  Collisions are erosive above and to the left of the solid line.  Collisions are increasingly accretionary for smaller $D$.  There is no accretion for impactors larger than 800~km except a narrow wedge near $sin (\theta) \approx 0.75$. }
\label{fig:mlr_moon}
\end{figure}

However, we do not expect the entire late veneer population to have been made up purely of 1000+~km-scale bodies.  Even if it were, collisional evolution would invariably create a population of fragments.  It is thus instructive to determine the outcome of impacts on the Moon's with somewhat smaller objects.  Figure~\ref{fig:mlr_moon} shows a map of the Moon's expected mass after experiencing a collision with a body with a range of sizes and at a range of impact angles, assuming $v_{rand} = 11.7 \, km \, s^{-1}$.  There is no net accretion for any impactors with $D > 800$~km except for a very narrow wedge of impact angles near $b \simeq 0.75$.  This critical value is anti-correlated with the impact speed: it decreases to $\sim 500$~km for $v_{rand} = 22 \, km \, s^{-1}$ and increases to 1000~km for $v_{rand} = 8.5 \, km \, s^{-1}$.  Impacts smaller than this velocity-dependent critical threshold are accretionary as long as they are not grazing so as to avoid the hit-and-run regime.  

Mercury also cannot accrete the largest impactors.  The median $v_{rand}$ for Mercury-impacting bodies was $23.9 \, km\, s^{-1}$, although the range of values spanned from $8$ to more than $50 \, km\,s^{-1}$.  For $v_{rand} =$ 15 / 24 / 35 $km\, s^{-1}$, Mercury cannot accrete bodies larger than $D =$1600 / 1100 / 850~km.  This size limit for accretion is clearly seen in Fig.~\ref{fig:limp_d}, where Mercury is not imparted any spin from impacts larger than $D \sim 2000$~km because of our assumption that only accretionary impacts impart spin angular momentum.  

Given that the Moon simply cannot accrete impactors larger than 500-1000~km, a top-heavy population of late veneer impactors must invariably generate a large Earth-to-Moon HSE abundance ratio, as observed.  If, however, the veneer population contained only smaller planetesimals then we would expect the Earth-to-Moon HSE abundance ratio to be closer to the ratio of their accretion cross sections, which can only match observations if the accretion cross section are significantly enhanced by gravitational focusing only feasible for very small ($R \sim 10$~m) planetesimals~\citep{schlichting12}.  ~\cite{bottke10}'s argument is that the Earth and Moon could have been impacted by different-sized objects if the veneer size distribution were shallow and extended to large objects.  However, Bottke et al. effectively underestimated the HSE abundance ratios they calculated -- as well as the success rate of their simulated populations -- by not accounting for non-accretionary or erosive collisions on the Moon.  

Any veneer population that includes impactors larger than 1000 km would naturally produce a large Earth-to-Moon HSE abundance ratio.  The Earth was the primary target of the late veneer impactors (along with Venus), while the Moon simply could not accrete any large bodies.  The HSEs in Earth's mantle could have been delivered by a number of 1000~km bodies or a single $\sim 3000$~km or larger one.  The Moon could not have participated in this accretion.  Both the Earth and Moon would have concurrently accreted $D \lesssim 800$~km bodies, with a ratio of roughly 25.  Thus, veneer impactors with $D \le 500-1000$~km could have had virtually any size distribution, including that of the present-day asteroid belt, as long as the population included a tail of $D \ge 1000$~km impactors.  However, the mass in these smaller ($D \le 500-1000$~km) veneer bodies is constrained to have been small enough that the Earth did not accrete more than $0.015 \mearth$ in total.  

For our argument to be correct, Earth must have accreted at least 90\% of its veneer mass from $D \ge 1000$~km impactors.  The Moon of course accreted no bodies this large, although it is possible that it was eroded during this phase (see discussion in Section 8).  The remaining 10\% of Earth's veneer could have come from a population of smaller $D \lesssim 1000$~km bodies that the Moon would also have accreted in roughly 1:25 proportion.  If more than 10\% of Earth's veneer came from small bodies then the Earth/Moon HSE mass ratio would drop below the lower measured limit of $\sim 200$.  Given that a late veneer with a total mass of $\sim 0.05 \mearth$ does a good job of reproducing the Earth's actual HSE abundance, this implies that the late veneer contained less than $5 \times 10^{-3} \mearth$ in $D \lesssim 1000$~km impactors.  While this is small compared with the planets' masses, it is nonetheless five times the total mass in the present-day asteroid belt.

\section{Is the stochastic late veneer hypothesis viable?}

At face value, our results are broadly consistent with the stochastic late veneer model of \cite{bottke10}.  However, there are consequences of our simulations, as well as assumptions of the model and connections with measured constraints, that warrant further discussion.  

First, it is important to point out that the start of the late veneer phase was different for each planet.  HSEs accumulate in the mantle once a planet's core becomes decoupled from the surface, after the last impact large enough to cause differentiation~\citep[e.g.,][]{tonks92}.  The last giant impact certainly occurred at a different time for each planet.  Geochemical constraints place this time at roughly 50-100 Myr for Earth~\citep{kleine09} and before $\sim 5$~Myr for Mars~\citep{dauphas11}.  HSE accumulation in the Martian mantle had thus been ongoing for 50 Myr or more before it began on Earth, making the duration of the late veneer phase 10-20\% longer for Mars than for Earth.  The fact that Earth and Mars have similar HSE concentrations in their mantles~\citep{walker09,brandon12} suggests that Mars did indeed accrete a modest surplus of veneer compared with Earth.  If the late veneer were determined purely by the gravitational cross sections $C$ (Eq. 6), then by using median values of $v_{rand}$ of $11.7 \, km \, s^{-1}$ for Earth and $8.3 \, km \, s^{-1}$ for Mars, then the Earth/Mars concentration ratio should be $\sim 0.6$, assuming that the mantle samples are a good representation of the bulk mantles.   Mars thus accreted more than its gravitational share in late veneer, presumably during the interval between its core closure at a few Myr and Earth's core closure after 50-100 Myr.   Given that the terrestrial planets started our simulations fully-formed, we might expect the Earth-to-Mars HSE abundance ratio to be somewhat underestimated in the simulations, which is indeed the case.  

The difference in the timing of core closure on Mars and Earth, combined with their HSE abundances, may offer a glimpse into the prevailing conditions in the early inner Solar System.  By the arguments presented above, we think that Mars accreted about about 40\% of its HSEs between its core closure and Earth's core closure $\sim 50$ Myr later.  This amounts to $1-4 \times 10^{-4} \mearth$ of mass accreted by Mars (the range assumes that Earth's late veneer was $0.003-0.01 \mearth$).  The young Earth was of course also accreting during this period.  With the median impact speeds from our late veneer simulations, the Earth's gravitational cross section $C$ is about 6.3 times larger than Mars'.  Earth should thus have accreted up to a few $\times 10^{-3}\mearth$ during the time between Mars' and Earth's core closures, i.e., between roughly 5 and 50 Myr after the start of planet formation.  However, it is thought that Earth accreted most or at least a significant fraction of its mass during this period~\citep[e.g.][]{chambers01,raymond06b,raymond09c,obrien06}.  Even if Earth only accreted 10\% of its final mass during this interval, this represents 40-160 times more than that expected from the arguments presented above.  How, then, did Mars accrete {\em so little} veneer during this time?  This is essentially the same issue as the ``small Mars'' problem~\citep{wetherill91,raymond09c}.  One possibility is that Mars's feeding zone was barren and contained far less mass than Earth's.  This is precisely the situation if most of the terrestrial planets' mass was concentrated in a belt, and Mars was scattered past the edge of the belt~\citep{hansen09}.  Mars' growth was effectively stunted by starvation in a low-density region beyond Earth and Venus.  The Grand Tack model proposes that the outer edge of the belt exists because it was truncated by Jupiter's gas-driven migration~\citep{walsh11}.   

Second, perhaps the most striking aspect of the stochastic late veneer model is the top-heavy distribution of veneer impactors.  These bodies are presumably the leftovers of terrestrial planet formation and thus were formed several millions of years before ``time zero'' in our simulations.  

The question thus arises of whether the planetesimal population that survived terrestrial accretion should realistically be dominated by $\gtrsim 1000$~km objects.  The initial size distribution of planetesimals is uncertain. Models of planetesimal formation by turbulent concentration can produce planetesimals as large as 100-1000km in diameter~\citep{johansen07,cuzzi08}. However, the size distribution of the asteroid belt can be reproduced by the collisional evolution of planetesimals that were initially hundreds of km in size~\citep{morby09b} or by coagulation of planetesimals that were initially only about 100m in size~\citep{weidenschilling11}. By the end of terrestrial planet formation, all planetesimals must have undergone some growth and collisional evolution so it is unclear how the veneer size distribution relates to the initial planetesimal size distribution. The size distribution of the planetesimals delivering the late veneer is therefore unknown.  If the planetesimal size distribution had a differential power law index of q=1-2 then the collisional cross section would have been dominated by the largest bodies.  The collision rate $f_{coll}$ between large bodies -- the collisions presumably required to destroy them -- can be very simply estimated as follows: 
\begin{equation}
f_{coll} \sim \frac{\Sigma \Omega}{\rho R},
\end{equation}
where $\Sigma$ is the surface density of bodies (taken to be simply $0.05 \mearth / \pi (1 \rm{AU})^2$), $\Omega$ is the Keplerian frequency at 1 AU, $\rho$ is the physical density of impactors assumed to be $3 \, g\, cm^{-3}$, and $R$ is the impactor radius.  The typical time between collisions is simply $t_{coll} = 1/f_{coll}$.  For $R$ = 500~km this simple estimate yields $t_{coll}$ = 56 Myr.  N-body simulations of the giant impact phase in the terrestrial planet region find time scales for the giant impact phase of 10-100~Myr~\citep{chambers01,raymond06b}. Since the giant impact and collisional timescales are comparable for a population dominated by large bodies, it is plausible that there was population of late veneer bodies dominated by large 1000~km sized impactors.

A third concern with the stochastic late veneer model is that, at first glance, our results appear to suggest that if the Earth's veneer came from large ($D>1000$~km) impactors then the Moon should have been significantly eroded.  Figure 3  shows that if the Moon were struck by the same impactors as Earth with the same random velocities, the vast majority of impacts would be in the catastrophic erosive regime.  Indeed, Figure 5 shows that at the relevant impact velocities the Moon is simply unable to accrete objects larger than $D \approx 800$~km.  

This discrepancy can be resolved by small number statistics.  Our simulations did not include objects smaller than 1000~km, and the $0.05 \mearth$ in veneer impactors needed to reproduce Earth's HSE abundance was contained in between 6 and 73 objects, depending on the size range of the bodies (i.e., the size of the largest bodies $D_{max}$) and the random number seed.  In the simulations Earth underwent between zero and 20 veneer impacts with a median of just 5 collisions.  At the relevant velocities the Earth's collisional cross section is about 25 times larger than the Moon's (see Section 7).  Given that this is larger than the maximum number of impacts on Earth, our simulations are consistent with the Moon not undergoing a single impact with a $D>1000$~km body.  

Nonetheless, is it conceivable that the Moon could have formed larger than it is and been eroded by a large veneer impact?  This possibility is not ruled out by our simulations; it is dynamically plausible that the Moon formed with twice its current mass and was later massively eroded by a very large veneer impact.  There exists a model capable of producing a Moon more massive than the actual one via a giant impact between the proto-Earth and a similar-mass embryo~\citep{canup12}.  Although this model does produce a Moon with the correct Oxygen isotope abundance, it may be inconsistent with geochemical evidence that the Earth's mantle was not well mixed even at very early times~\citep{touboul12,tucker13}.  

Given that the arguments against the formation of a larger Moon are relatively indirect, the possibility that the Moon formed modestly larger than it is today appears to be a viable scenario, although we cannot place strong constraints on how much larger with the current simulations.  While most giant impacts on the Moon would be erosive, more than two-thirds of $D = 1000-4000$~km impacts would erode the Moon by less than 25\%.  Among $D < 2000$~km objects, only 22\% of impacts erode the Moon by more than 25\%.  The formation of a Moon up to, say, 25\% more massive than the current one remains a dynamically plausible and interesting scenario.  It may be interesting to search for geochemical or geological signatures of lunar erosion. 

Finally, we note that the threshold for core formation is poorly understood.  We have inherently assumed that all veneer impactors are below the threshold for differentiation regardless of planet size, impactor size and velocity.  In practice, we suspect that an impactor's must be smaller than the thickness of the planet's mantle to remain in the mantle and avoid sequestering at least a portion of the planet's HSEs in the core.  However, further work is needed in this area to achieve an understanding of the fate of HSEs in large impacts.  A sufficiently energetic impact should produce a magma ocean of a given depth, and some Earth impacts in our simulations have impact velocities in excess of $20 \, km \, s^{-1}$.  The most energetic impacts in the simulations are only a factor of 5-20 smaller than the impact energy in recent simulations of lunar formation~\citep{cuk12,canup12}.  In the context of our simulations and collisional model we have assumed that these impacts deposited HSEs in Earth's mantle, although further studies of magma ocean production and partial differentiation after large impacts would help in the interpretation of our simulations.  

\section{Conclusions}

In this paper we used N-body simulations to test the stochastic late veneer model proposed by~\cite{bottke10}.  We found that the model is viable with some constraints.  First, the total mass in the veneer population cannot have been much higher than $\sim 0.05 \mearth$.  If it were there should be a higher concentration of HSEs in Earth's crust.  Second, the terrestrial planets' pre-veneer orbits were as calm or calmer than their present ones.  Given that the terrestrial planets' angular momentum deficit $AMD$ immediately before the late heavy bombardment must have been less than $0.7 \, AMD_{now}$~\citep{brasser13}, we find a drastically higher success rate in simulations with initial $AMD_0 \lesssim 0.5 \, AMD_{now}$.  For somewhat smaller veneer impactors ($D_{max} = 2000$~km), the initial $AMD$ could have been as high as $AMD_{now}$.  Such low $AMD$ values are consistent with recent simulations of terrestrial planet formation, in particular if the terrestrial planets formed from a narrow annulus of material~\citep{hansen09,walsh11}.  
 
~\cite{bottke10} reproduced the large Earth/Moon HSE abundance ratio with a shallow veneer size distribution extending to large ($D \ge 2000$~km) impactors.  Using the model for the outcome of gravity-dominated collisions developed by~\cite{leinhardt12}, we showed that the Moon simply cannot accrete objects larger than $D \approx 800$~km (Fig. 5; note that this critical size is velocity dependent and extends from $D \approx 500-1000$~km for relevant parameters).  Rather, the Moon would be modestly eroded by $D$=1000-4000~km impactors, typically by 10-20\%.  The large Earth/Moon HSE abundance ratio is naturally reproduced by any top-heavy impacting population simply because the Moon does not accrete in this regime while Earth does.  All that is required is that $\sim10$\% or less of the veneer's total mass -- $\sim 5 \times 10^{-3} \mearth$ -- was initially in small ($D \lesssim 500-1000$~km) bodies, as these are accreted by both the Moon and Earth in roughly 1:25 proportion.  Of course, a later large impact could strip some HSEs from the Moon.  

A stochastic late veneer has dramatic consequences for some of the planets.  While this is an accretionary phase for Venus, Earth and Mars, it is an erosive one for Mercury and the Moon.  Mercury typically lost a few percent of its mass via high-speed collisions and in a considerable fraction of cases (17\%) an erosive collision would have removed more than a quarter of its total mass.  Although we did not track the long-term evolution of the impact debris, such erosive impacts may explain Mercury's large iron mass fraction~\citep{benz88,benz07}.  Any impact of a $D \ge$ 500-1000~km body would have eroded the Moon (see discussion in Section 8).  Likewise, almost any impactor larger than $D =$ 850-1600~km would have eroded Mercury.  It is worth considering the possibility that the primordial Moon was more massive than the current one, perhaps by up to 25\%. 

A stochastic late veneer would have had a strong effect on the spins of Mercury and Venus.  Veneer impacts delivered 1-1000 times those planets' current spin angular momenta (Fig. 4), suggesting that their primordial spin rates were probably much faster than their current ones, which have been altered by tidal friction~\citep{correia01,correia09}.

\vskip 1in
{\it Acknowledgments.} 
We thank Sarah Stewart, an anonymous second referee, and editor Alessandro Morbidelli for their constructive criticism of the paper.  We thank Nate Kaib for the original version of the code to extract collision information from Mercury runfiles.  S.N.R. acknowledges support from the CNRS's PNP program and NASA Astrobiology Institute's Virtual Planetary Laboratory lead team.  For H.S. support for this work was provided  by NASA through Hubble Fellowship Grant \# HST-HF-51281.01-A awarded by the Space Telescope Science Institute, which is operated by the Association of Universities for Research in Astronomy, Inc., for NASA, under contact NAS 5-26555.


\begin{thebibliography}{}

\bibitem[{Agnor} {\em et~al.}(1999){Agnor}, {Canup}, and {Levison}]{agnor99}
{Agnor}, C.~B., R.~M. {Canup},\ and H.~F. {Levison} 1999.
\newblock {On the Character and Consequences of Large Impacts in the Late Stage
  of Terrestrial Planet Formation}.
\newblock {\em \icarus\/}~{\bf 142}, 219--237.

\bibitem[{Agnor} and {Lin}(2012){Agnor} and {Lin}]{agnor12}
{Agnor}, C.~B.,\ and D.~N.~C. {Lin} 2012.
\newblock {On the Migration of Jupiter and Saturn: Constraints from Linear
  Models of Secular Resonant Coupling with the Terrestrial Planets}.
\newblock {\em \apj\/}~{\bf 745}, 143.

\bibitem[{Asphaug}(2010){Asphaug}]{asphaug10}
{Asphaug}, E. 2010.
\newblock {Similar-sized collisions and the diversity of planets}.
\newblock {\em Chemie der Erde / Geochemistry\/}~{\bf 70}, 199--219.

\bibitem[{Asphaug} {\em et~al.}(2006){Asphaug}, {Agnor}, and
  {Williams}]{asphaug06}
{Asphaug}, E., C.~B. {Agnor},\ and Q.~{Williams} 2006.
\newblock {Hit-and-run planetary collisions}.
\newblock {\em \nat\/}~{\bf 439}, 155--160.

\bibitem[{Batygin} and {Brown}(2010){Batygin} and {Brown}]{batygin10}
{Batygin}, K.,\ and M.~E. {Brown} 2010.
\newblock {Early Dynamical Evolution of the Solar System: Pinning Down the
  Initial Conditions of the Nice Model}.
\newblock {\em \apj\/}~{\bf 716}, 1323--1331.

\bibitem[{Benz} {\em et~al.}(2007){Benz}, {Anic}, {Horner}, and
  {Whitby}]{benz07}
{Benz}, W., A.~{Anic}, J.~{Horner},\ and J.~A. {Whitby} 2007.
\newblock {The Origin of Mercury}.
\newblock {\em \ssr\/}~{\bf 132}, 189--202.

\bibitem[{Benz} {\em et~al.}(1986){Benz}, {Slattery}, and {Cameron}]{benz86}
{Benz}, W., W.~L. {Slattery},\ and A.~G.~W. {Cameron} 1986.
\newblock {The origin of the moon and the single-impact hypothesis. I}.
\newblock {\em \icarus\/}~{\bf 66}, 515--535.

\bibitem[{Benz} {\em et~al.}(1988){Benz}, {Slattery}, and {Cameron}]{benz88}
{Benz}, W., W.~L. {Slattery},\ and A.~G.~W. {Cameron} 1988.
\newblock {Collisional stripping of Mercury's mantle}.
\newblock {\em \icarus\/}~{\bf 74}, 516--528.

\bibitem[{Bottke} {\em et~al.}(2010){Bottke}, {Walker}, {Day}, {Nesvorny}, and
  {Elkins-Tanton}]{bottke10}
{Bottke}, W.~F., R.~J. {Walker}, J.~M.~D. {Day}, D.~{Nesvorny},\ and
  L.~{Elkins-Tanton} 2010.
\newblock {Stochastic Late Accretion to Earth, the Moon, and Mars}.
\newblock {\em Science\/}~{\bf 330}, 1527--.

\bibitem[{Brandon} {\em et~al.}(2012){Brandon}, {Puchtel}, {Walker}, {Day},
  {Irving}, and {Taylor}]{brandon12}
{Brandon}, A.~D., I.~S. {Puchtel}, R.~J. {Walker}, J.~M.~D. {Day}, A.~J.
  {Irving},\ and L.~A. {Taylor} 2012.
\newblock {Evolution of the martian mantle inferred from the
  $^{187}$Re-$^{187}$Os isotope and highly siderophile element abundance
  systematics of shergottite meteorites}.
\newblock {\em \gca\/}~{\bf 76}, 206--235.

\bibitem[{Brasser} {\em et~al.}(2009){Brasser}, {Morbidelli}, {Gomes},
  {Tsiganis}, and {Levison}]{brasser09}
{Brasser}, R., A.~{Morbidelli}, R.~{Gomes}, K.~{Tsiganis},\ and H.~F. {Levison}
  2009.
\newblock {Constructing the secular architecture of the solar system II: the
  terrestrial planets}.
\newblock {\em \aap\/}~{\bf 507}, 1053--1065.

\bibitem[{Brasser} {\em et~al.}(2013){Brasser}, {Walsh}, and
  {Nesvorny}]{brasser13}
{Brasser}, R., K.~{Walsh},\ and D.~{Nesvorny} 2013.
\newblock {Constraining the primordial orbits of the Terrestrial Planets}.
\newblock {\em ArXiv e-prints\/}.

\bibitem[{Canup}(2012){Canup}]{canup12}
{Canup}, R.~M. 2012.
\newblock {Forming a Moon with an Earth-like Composition via a Giant Impact}.
\newblock {\em Science\/}~{\bf 338}, 1052--.

\bibitem[{Canup} and {Asphaug}(2001){Canup} and {Asphaug}]{canup01}
{Canup}, R.~M.,\ and E.~{Asphaug} 2001.
\newblock {Origin of the Moon in a giant impact near the end of the Earth's
  formation}.
\newblock {\em \nat\/}~{\bf 412}, 708--712.

\bibitem[{Chambers}(1999){Chambers}]{chambers99}
{Chambers}, J.~E. 1999.
\newblock {A hybrid symplectic integrator that permits close encounters between
  massive bodies}.
\newblock {\em \mnras\/}~{\bf 304}, 793--799.

\bibitem[{Chambers}(2001){Chambers}]{chambers01}
{Chambers}, J.~E. 2001.
\newblock {Making More Terrestrial Planets}.
\newblock {\em Icarus\/}~{\bf 152}, 205--224.

\bibitem[{Chambers}(2013){Chambers}]{chambers13}
{Chambers}, J.~E. 2013.
\newblock {Late-stage planetary accretion including hit-and-run collisions and
  fragmentation}.
\newblock {\em \icarus\/}~{\bf 224}, 43--56.

\bibitem[{Chambers} and {Wetherill}(1998){Chambers} and
  {Wetherill}]{chambers98}
{Chambers}, J.~E.,\ and G.~W. {Wetherill} 1998.
\newblock {Making the Terrestrial Planets: N-Body Integrations of Planetary
  Embryos in Three Dimensions}.
\newblock {\em \icarus\/}~{\bf 136}, 304--327.

\bibitem[{Correia} and {Laskar}(2001){Correia} and {Laskar}]{correia01}
{Correia}, A.~C.~M.,\ and J.~{Laskar} 2001.
\newblock {The four final rotation states of Venus}.
\newblock {\em \nat\/}~{\bf 411}, 767--770.

\bibitem[{Correia} and {Laskar}(2009){Correia} and {Laskar}]{correia09}
{Correia}, A.~C.~M.,\ and J.~{Laskar} 2009.
\newblock {Mercury's capture into the 3/2 spin-orbit resonance including the
  effect of core-mantle friction}.
\newblock {\em \icarus\/}~{\bf 201}, 1--11.

\bibitem[{{\'C}uk} and {Stewart}(2012){{\'C}uk} and {Stewart}]{cuk12}
{{\'C}uk}, M.,\ and S.~T. {Stewart} 2012.
\newblock {Making the Moon from a Fast-Spinning Earth: A Giant Impact Followed
  by Resonant Despinning}.
\newblock {\em Science\/}~{\bf 338}, 1047--.

\bibitem[{Cuzzi} {\em et~al.}(2008){Cuzzi}, {Hogan}, and {Shariff}]{cuzzi08}
{Cuzzi}, J.~N., R.~C. {Hogan},\ and K.~{Shariff} 2008.
\newblock {Toward Planetesimals: Dense Chondrule Clumps in the Protoplanetary
  Nebula}.
\newblock {\em \apj\/}~{\bf 687}, 1432--1447.

\bibitem[{Dauphas} and {Pourmand}(2011){Dauphas} and {Pourmand}]{dauphas11}
{Dauphas}, N.,\ and A.~{Pourmand} 2011.
\newblock {Hf-W-Th evidence for rapid growth of Mars and its status as a
  planetary embryo}.
\newblock {\em \nat\/}~{\bf 473}, 489--492.

\bibitem[{Day} {\em et~al.}(2007){Day}, {Pearson}, and {Taylor}]{day07}
{Day}, J.~M.~D., D.~G. {Pearson},\ and L.~A. {Taylor} 2007.
\newblock {Highly Siderophile Element Constraints on Accretion and
  Differentiation of the Earth-Moon System}.
\newblock {\em Science\/}~{\bf 315}, 217--.

\bibitem[{Dones} and {Tremaine}(1993){Dones} and {Tremaine}]{dones93}
{Dones}, L.,\ and S.~{Tremaine} 1993.
\newblock {On the origin of planetary spins}.
\newblock {\em \icarus\/}~{\bf 103}, 67--92.

\bibitem[{Genda} {\em et~al.}(2012){Genda}, {Kokubo}, and {Ida}]{genda12}
{Genda}, H., E.~{Kokubo},\ and S.~{Ida} 2012.
\newblock {Merging Criteria for Giant Impacts of Protoplanets}.
\newblock {\em \apj\/}~{\bf 744}, 137.

\bibitem[{Goldreich}(1966){Goldreich}]{goldreich66}
{Goldreich}, P. 1966.
\newblock {History of the Lunar Orbit}.
\newblock {\em Reviews of Geophysics and Space Physics\/}~{\bf 4}, 411--439.

\bibitem[{Hansen}(2009){Hansen}]{hansen09}
{Hansen}, B.~M.~S. 2009.
\newblock {Formation of the Terrestrial Planets from a Narrow Annulus}.
\newblock {\em \apj\/}~{\bf 703}, 1131--1140.

\bibitem[{Jackson} and {Wyatt}(2012){Jackson} and {Wyatt}]{jackson12}
{Jackson}, A.~P.,\ and M.~C. {Wyatt} 2012.
\newblock {Debris from terrestrial planet formation: the Moon-forming
  collision}.
\newblock {\em \mnras\/}~{\bf 425}, 657--679.

\bibitem[{Johansen} and {Lacerda}(2010){Johansen} and {Lacerda}]{johansen10}
{Johansen}, A.,\ and P.~{Lacerda} 2010.
\newblock {Prograde rotation of protoplanets by accretion of pebbles in a
  gaseous environment}.
\newblock {\em \mnras\/}~{\bf 404}, 475--485.

\bibitem[{Johansen} {\em et~al.}(2007){Johansen}, {Oishi}, {Mac Low}, {Klahr},
  {Henning}, and {Youdin}]{johansen07}
{Johansen}, A., J.~S. {Oishi}, M.-M. {Mac Low}, H.~{Klahr}, T.~{Henning},\ and
  A.~{Youdin} 2007.
\newblock {Rapid planetesimal formation in turbulent circumstellar disks}.
\newblock {\em \nat\/}~{\bf 448}, 1022--1025.

\bibitem[{Kaula}(1964){Kaula}]{kaula64}
{Kaula}, W.~M. 1964.
\newblock {Tidal Dissipation by Solid Friction and the Resulting Orbital
  Evolution}.
\newblock {\em Reviews of Geophysics and Space Physics\/}~{\bf 2}, 661--685.

\bibitem[{Kimura} {\em et~al.}(1974){Kimura}, {Lewis}, and {Anders}]{kimura74}
{Kimura}, K., R.~S. {Lewis},\ and E.~{Anders} 1974.
\newblock {Distribution of gold and rhenium between nickel-iron and silicate
  melts: implications for the abundance of siderophile elements on the Earth
  and Moon}.
\newblock {\em \gca\/}~{\bf 38}, 683--701.

\bibitem[{Kleine} {\em et~al.}(2009){Kleine}, {Touboul}, {Bourdon}, {Nimmo},
  {Mezger}, {Palme}, {Jacobsen}, {Yin}, and {Halliday}]{kleine09}
{Kleine}, T., M.~{Touboul}, B.~{Bourdon}, F.~{Nimmo}, K.~{Mezger}, H.~{Palme},
  S.~B. {Jacobsen}, Q.-Z. {Yin},\ and A.~N. {Halliday} 2009.
\newblock {Hf-W chronology of the accretion and early evolution of asteroids
  and terrestrial planets}.
\newblock {\em \gca\/}~{\bf 73}, 5150--5188.

\bibitem[{Kokubo} and {Ida}(2007){Kokubo} and {Ida}]{kokubo07}
{Kokubo}, E.,\ and S.~{Ida} 2007.
\newblock {Formation of Terrestrial Planets from Protoplanets. II. Statistics
  of Planetary Spin}.
\newblock {\em \apj\/}~{\bf 671}, 2082--2090.

\bibitem[{Laskar}(1997){Laskar}]{laskar97}
{Laskar}, J. 1997.
\newblock {Large scale chaos and the spacing of the inner planets.}
\newblock {\em \aap\/}~{\bf 317}, L75--L78.

\bibitem[{Laskar} {\em et~al.}(1993){Laskar}, {Joutel}, and {Boudin}]{laskar93}
{Laskar}, J., F.~{Joutel},\ and F.~{Boudin} 1993.
\newblock {Orbital, precessional, and insolation quantities for the Earth from
  -20 MYR to +10 MYR}.
\newblock {\em \aap\/}~{\bf 270}, 522--533.

\bibitem[{Leinhardt} and {Stewart}(2012){Leinhardt} and {Stewart}]{leinhardt12}
{Leinhardt}, Z.~M.,\ and S.~T. {Stewart} 2012.
\newblock {Collisions between Gravity-dominated Bodies. I. Outcome Regimes and
  Scaling Laws}.
\newblock {\em \apj\/}~{\bf 745}, 79.

\bibitem[{Love} and {Ahrens}(1997){Love} and {Ahrens}]{love97}
{Love}, S.~G.,\ and T.~J. {Ahrens} 1997.
\newblock {Origin of asteroid rotation rates in catastrophic impacts}.
\newblock {\em \nat\/}~{\bf 386}, 154--156.

\bibitem[{Masset} and {Snellgrove}(2001){Masset} and {Snellgrove}]{masset01}
{Masset}, F.,\ and M.~{Snellgrove} 2001.
\newblock {Reversing type II migration: resonance trapping of a lighter giant
  protoplanet}.
\newblock {\em \mnras\/}~{\bf 320}, L55--L59.

\bibitem[{Morbidelli} {\em et~al.}(2009){Morbidelli}, {Bottke}, {Nesvorn{\'y}},
  and {Levison}]{morby09b}
{Morbidelli}, A., W.~F. {Bottke}, D.~{Nesvorn{\'y}},\ and H.~F. {Levison} 2009.
\newblock {Asteroids were born big}.
\newblock {\em \icarus\/}~{\bf 204}, 558--573.

\bibitem[{Morbidelli} {\em et~al.}(2010){Morbidelli}, {Brasser}, {Gomes},
  {Levison}, and {Tsiganis}]{morby10}
{Morbidelli}, A., R.~{Brasser}, R.~{Gomes}, H.~F. {Levison},\ and K.~{Tsiganis}
  2010.
\newblock {Evidence from the Asteroid Belt for a Violent Past Evolution of
  Jupiter's Orbit}.
\newblock {\em \aj\/}~{\bf 140}, 1391--1401.

\bibitem[{Morbidelli} {\em et~al.}(2009){Morbidelli}, {Brasser}, {Tsiganis},
  {Gomes}, and {Levison}]{morby09a}
{Morbidelli}, A., R.~{Brasser}, K.~{Tsiganis}, R.~{Gomes},\ and H.~F. {Levison}
  2009.
\newblock {Constructing the secular architecture of the solar system. I. The
  giant planets}.
\newblock {\em \aap\/}~{\bf 507}, 1041--1052.

\bibitem[{Morbidelli} {\em et~al.}(2007){Morbidelli}, {Tsiganis}, {Crida},
  {Levison}, and {Gomes}]{morby07b}
{Morbidelli}, A., K.~{Tsiganis}, A.~{Crida}, H.~F. {Levison},\ and R.~{Gomes}
  2007.
\newblock {Dynamics of the Giant Planets of the Solar System in the Gaseous
  Protoplanetary Disk and Their Relationship to the Current Orbital
  Architecture}.
\newblock {\em \aj\/}~{\bf 134}, 1790--1798.

\bibitem[{Nesvorn{\'y}} and {Morbidelli}(2012){Nesvorn{\'y}} and
  {Morbidelli}]{nesvorny12}
{Nesvorn{\'y}}, D.,\ and A.~{Morbidelli} 2012.
\newblock {Statistical Study of the Early Solar System's Instability with Four,
  Five, and Six Giant Planets}.
\newblock {\em \aj\/}~{\bf 144}, 117.

\bibitem[{O'Brien} {\em et~al.}(2006){O'Brien}, {Morbidelli}, and
  {Levison}]{obrien06}
{O'Brien}, D.~P., A.~{Morbidelli},\ and H.~F. {Levison} 2006.
\newblock {Terrestrial planet formation with strong dynamical friction}.
\newblock {\em \icarus\/}~{\bf 184}, 39--58.

\bibitem[{Pierens} and {Raymond}(2011){Pierens} and {Raymond}]{pierens11}
{Pierens}, A.,\ and S.~N. {Raymond} 2011.
\newblock {Two phase, inward-then-outward migration of Jupiter and Saturn in
  the gaseous solar nebula}.
\newblock {\em \aap\/}~{\bf 533}, A131.

\bibitem[{Quinn} {\em et~al.}(1991){Quinn}, {Tremaine}, and {Duncan}]{quinn91}
{Quinn}, T.~R., S.~{Tremaine},\ and M.~{Duncan} 1991.
\newblock {A three million year integration of the earth's orbit}.
\newblock {\em \aj\/}~{\bf 101}, 2287--2305.

\bibitem[{Raymond} {\em et~al.}(2011){Raymond}, {Armitage},
  {Moro-Mart{\'{\i}}n}, {Booth}, {Wyatt}, {Armstrong}, {Mandell}, {Selsis}, and
  {West}]{raymond11}
{Raymond}, S.~N., P.~J. {Armitage}, A.~{Moro-Mart{\'{\i}}n}, M.~{Booth}, M.~C.
  {Wyatt}, J.~C. {Armstrong}, A.~M. {Mandell}, F.~{Selsis},\ and A.~A. {West}
  2011.
\newblock {Debris disks as signposts of terrestrial planet formation}.
\newblock {\em \aap\/}~{\bf 530}, A62.

\bibitem[{Raymond} {\em et~al.}(2009){Raymond}, {O'Brien}, {Morbidelli}, and
  {Kaib}]{raymond09c}
{Raymond}, S.~N., D.~P. {O'Brien}, A.~{Morbidelli},\ and N.~A. {Kaib} 2009.
\newblock {Building the terrestrial planets: Constrained accretion in the inner
  Solar System}.
\newblock {\em Icarus\/}~{\bf 203}, 644--662.

\bibitem[{Raymond} {\em et~al.}(2004){Raymond}, {Quinn}, and
  {Lunine}]{raymond04}
{Raymond}, S.~N., T.~{Quinn},\ and J.~I. {Lunine} 2004.
\newblock {Making other earths: dynamical simulations of terrestrial planet
  formation and water delivery}.
\newblock {\em Icarus\/}~{\bf 168}, 1--17.

\bibitem[{Raymond} {\em et~al.}(2006){Raymond}, {Quinn}, and
  {Lunine}]{raymond06b}
{Raymond}, S.~N., T.~{Quinn},\ and J.~I. {Lunine} 2006.
\newblock {High-resolution simulations of the final assembly of Earth-like
  planets I. Terrestrial accretion and dynamics}.
\newblock {\em \icarus\/}~{\bf 183}, 265--282.

\bibitem[{Rubie} {\em et~al.}(2003){Rubie}, {Melosh}, {Reid}, {Liebske}, and
  {Righter}]{rubie03}
{Rubie}, D.~C., H.~J. {Melosh}, J.~E. {Reid}, C.~{Liebske},\ and K.~{Righter}
  2003.
\newblock {Mechanisms of metal-silicate equilibration in the terrestrial magma
  ocean}.
\newblock {\em Earth and Planetary Science Letters\/}~{\bf 205}, 239--255.

\bibitem[{Safronov}(1969){Safronov}]{safronov69}
{Safronov}, V.~S. 1969.
\newblock {\em {Evoliutsiia doplanetnogo oblaka.}}

\bibitem[{Schlichting} and {Sari}(2007){Schlichting} and {Sari}]{schlichting07}
{Schlichting}, H.~E.,\ and R.~{Sari} 2007.
\newblock {The Effect of Semicollisional Accretion on Planetary Spins}.
\newblock {\em \apj\/}~{\bf 658}, 593--597.

\bibitem[{Schlichting} {\em et~al.}(2012){Schlichting}, {Warren}, and
  {Yin}]{schlichting12}
{Schlichting}, H.~E., P.~H. {Warren},\ and Q.-Z. {Yin} 2012.
\newblock {The Last Stages of Terrestrial Planet Formation: Dynamical Friction
  and the Late Veneer}.
\newblock {\em \apj\/}~{\bf 752}, 8.

\bibitem[{Tera} {\em et~al.}(1974){Tera}, {Papanastassiou}, and
  {Wasserburg}]{tera74}
{Tera}, F., D.~A. {Papanastassiou},\ and G.~J. {Wasserburg} 1974.
\newblock {Isotopic evidence for a terminal lunar cataclysm}.
\newblock {\em Earth and Planetary Science Letters\/}~{\bf 22}, 1.

\bibitem[{Tonks} and {Melosh}(1992){Tonks} and {Melosh}]{tonks92}
{Tonks}, W.~B.,\ and H.~J. {Melosh} 1992.
\newblock {Core formation by giant impacts}.
\newblock {\em \icarus\/}~{\bf 100}, 326--346.

\bibitem[{Touboul} {\em et~al.}(2012){Touboul}, {Puchtel}, and
  {Walker}]{touboul12}
{Touboul}, M., I.~S. {Puchtel},\ and R.~J. {Walker} 2012.
\newblock {$^{182}$W Evidence for Long-Term Preservation of Early Mantle
  Differentiation Products}.
\newblock {\em Science\/}~{\bf 335}, 1065--.

\bibitem[{Tsiganis} {\em et~al.}(2005){Tsiganis}, {Gomes}, {Morbidelli}, and
  {Levison}]{tsiganis05}
{Tsiganis}, K., R.~{Gomes}, A.~{Morbidelli},\ and H.~F. {Levison} 2005.
\newblock {Origin of the orbital architecture of the giant planets of the Solar
  System}.
\newblock {\em \nat\/}~{\bf 435}, 459--461.

\bibitem[{Tucker} and {Mukhopadhyay}(2013){Tucker} and
  {Mukhopadhyay}]{tucker13}
{Tucker}, J.~M.,\ and S.~{Mukhopadhyay} 2013.
\newblock {Evidence for Multiple Giant Impacts and Magma Oceans from Mantle
  Noble Gases}.
\newblock {\em LPI Contributions\/}~{\bf 1719}, 2990.

\bibitem[{Walker}(2009){Walker}]{walker09}
{Walker}, R.~J. 2009.
\newblock {Highly siderophile elements in the Earth, Moon and Mars: Update and
  implications for planetary accretion and differentiation}.
\newblock {\em Chemie der Erde / Geochemistry\/}~{\bf 69}, 101--125.

\bibitem[{Walker} {\em et~al.}(2004){Walker}, {Horan}, {Shearer}, and
  {Papike}]{walker04}
{Walker}, R.~J., M.~F. {Horan}, C.~K. {Shearer},\ and J.~J. {Papike} 2004.
\newblock {Low abundances of highly siderophile elements in the lunar mantle:
  evidence for prolonged late accretion}.
\newblock {\em Earth and Planetary Science Letters\/}~{\bf 224}, 399--413.

\bibitem[{Walsh} {\em et~al.}(2011){Walsh}, {Morbidelli}, {Raymond}, {O'Brien},
  and {Mandell}]{walsh11}
{Walsh}, K.~J., A.~{Morbidelli}, S.~N. {Raymond}, D.~P. {O'Brien},\ and A.~M.
  {Mandell} 2011.
\newblock {A low mass for Mars from Jupiter's early gas-driven migration}.
\newblock {\em \nat\/}~{\bf 475}, 206--209.

\bibitem[{Weidenschilling}(2011){Weidenschilling}]{weidenschilling11}
{Weidenschilling}, S.~J. 2011.
\newblock {Initial sizes of planetesimals and accretion of the asteroids}.
\newblock {\em \icarus\/}~{\bf 214}, 671--684.

\bibitem[{Wetherill}(1975){Wetherill}]{wetherill75}
{Wetherill}, G.~W. 1975.
\newblock {Late heavy bombardment of the moon and terrestrial planets}.
\newblock In {\em Lunar and Planetary Science Conference Proceedings}, Volume~6
  of {\em Lunar and Planetary Science Conference Proceedings}, pp.\
  1539--1561.

\bibitem[{Wetherill}(1985){Wetherill}]{wetherill85}
{Wetherill}, G.~W. 1985.
\newblock {Occurrence of giant impacts during the growth of the terrestrial
  planets}.
\newblock {\em Science\/}~{\bf 228}, 877--879.

\bibitem[{Wetherill}(1991){Wetherill}]{wetherill91}
{Wetherill}, G.~W. 1991.
\newblock {Why Isn't Mars as Big as Earth?}
\newblock In {\em Lunar and Planetary Institute Science Conference Abstracts},
  Volume~22 of {\em Lunar and Planetary Institute Science Conference
  Abstracts}, pp.\  1495.

\bibitem[{Wieczorek} {\em et~al.}(2012){Wieczorek}, {Correia}, {Le Feuvre},
  {Laskar}, and {Rambaux}]{wieczorek12}
{Wieczorek}, M.~A., A.~C.~M. {Correia}, M.~{Le Feuvre}, J.~{Laskar},\ and
  N.~{Rambaux} 2012.
\newblock {Mercury's spin-orbit resonance explained by initial retrograde and
  subsequent synchronous rotation}.
\newblock {\em Nature Geoscience\/}~{\bf 5}, 18--21.

\bibitem[{Willbold} {\em et~al.}(2011){Willbold}, {Elliott}, and
  {Moorbath}]{willbold11}
{Willbold}, M., T.~{Elliott},\ and S.~{Moorbath} 2011.
\newblock {The tungsten isotopic composition of the Earth's mantle before the
  terminal bombardment}.
\newblock {\em \nat\/}~{\bf 477}, 195--198.

\end{thebibliography}

\end{document}